\documentclass[10pt,conference]{IEEEtran}
\IEEEoverridecommandlockouts
\usepackage{cite}
\usepackage{amsmath,amssymb,amsfonts}
\usepackage{algorithmic}
\usepackage{graphicx}
\usepackage{textcomp}
\usepackage[table]{xcolor}
\usepackage[ruled,linesnumbered]{algorithm2e}
\usepackage{subfigure}
\usepackage{multirow} 
\usepackage{booktabs}
\usepackage{url}

\usepackage{bbding}
\usepackage[final]{changes}

\definecolor{pos}{RGB}{225, 225, 225}

\usepackage{tcolorbox}
\definechangesauthor[name={Dcm}, color=red]{dcm.}
\def\BibTeX{{\rm B\kern-.05em{\sc i\kern-.025em b}\kern-.08em
    T\kern-.1667em\lower.7ex\hbox{E}\kern-.125emX}}
\begin{document}

\title{Walk the Talk: Is Your Log-based Software Reliability Maintenance System Really Reliable?\\
\thanks{\IEEEauthorrefmark{1} Corresponding Authors}
}

\author{
    \IEEEauthorblockN{
        Minghua He\IEEEauthorrefmark{2},
        Tong Jia\IEEEauthorrefmark{2}\IEEEauthorrefmark{3}\IEEEauthorrefmark{1},
        Chiming Duan\IEEEauthorrefmark{2},
        Pei Xiao\IEEEauthorrefmark{2}, \\
        Lingzhe Zhang\IEEEauthorrefmark{2},
        Kangjin Wang\IEEEauthorrefmark{4},
        Yifan Wu\IEEEauthorrefmark{2},
        Ying Li\IEEEauthorrefmark{2}\IEEEauthorrefmark{1},
        and Gang Huang\IEEEauthorrefmark{2}\IEEEauthorrefmark{3}
    }
    \IEEEauthorblockA
    {
    \IEEEauthorrefmark{2}Peking University, Beijing, China\\ 
    \IEEEauthorrefmark{3}National Key Laboratory of Data Space Technology and System, Beijing, China\\ 
    \IEEEauthorrefmark{4}Alibaba Group, Hangzhou, China \\
    \{hemh2120, duanchiming, xiaopei, zhang.lingzhe\}@stu.pku.edu.cn \\ 
    \{jia.tong, yifanwu, li.ying, hg\}@pku.edu.cn, kangjin.wkj@alibaba-inc.com
    }
}

% \and
% \and
% \and
% \and
% \and

\maketitle

\begin{abstract}
%\textbf{Todo}

Log-based software reliability maintenance systems are crucial for sustaining stable customer experience. However, existing deep learning-based methods represent a black box for service providers, making it impossible for providers to understand how these methods detect anomalies, thereby hindering trust and deployment in real production environments. To address this issue, this paper defines a trustworthiness metric—diagnostic faithfulness—for models to gain service providers' trust, based on surveys of SREs at a major cloud provider. We design two evaluation tasks: attention-based root cause localization and event perturbation. Empirical studies demonstrate that existing methods perform poorly in diagnostic faithfulness. Consequently, we propose FaithLog, a faithful log-based anomaly detection system, which achieves faithfulness through a carefully designed causality-guided attention mechanism and adversarial consistency learning. Evaluation results on two public datasets and one industrial dataset demonstrate that the proposed method achieves state-of-the-art performance in diagnostic faithfulness.
% Code is available: \url{https://reglog4ad.github.io/}

\end{abstract}


\section{Introduction}

Software systems are growing increasingly large and complex, with their composition and operational logic becoming more intricate, while being subject to more failures \cite{niu2023locating, lin2023edits, li2023elastic, luo2021ntam}. In the real world where availability is critical to customer experience, service providers have designed software reliability maintenance systems that aim to rapidly and efficiently manage anomalous events to enhance reliability \cite{jiang2011efficient, zhang2024failure}.

Logs record system operational states and serve as the primary observable data utilized by software reliability maintenance systems. The use of logs for detecting system anomalies has been extensively studied \cite{guo2024logformer,sun2025exploring,zhao2025few,tong2016approach,devops,yin2020improving,kim2020automatic,zhang2024metalog,jia2024hilogx}. Recent approaches focus on deep learning, particularly attention-based sequence models \cite{zhang2019robust, le2021log, li2022swisslog, plelog, llmelog}. These methods employ sequential neural networks to extract system operational patterns from logs, then utilize attention mechanisms \cite{vaswani2017attention} to assign differential weights to distinct system events, and finally integrate diverse events to detect system anomalies. This strategy demonstrates the capability to learn historical operational patterns of systems, thereby enabling anomaly detection and has been proven effective in maintaining software system reliability \cite{lee2023eadro, midlog, le2024prelog, afalog}.

\begin{figure}[htbp]
\centerline{
\includegraphics[width=0.9\linewidth]{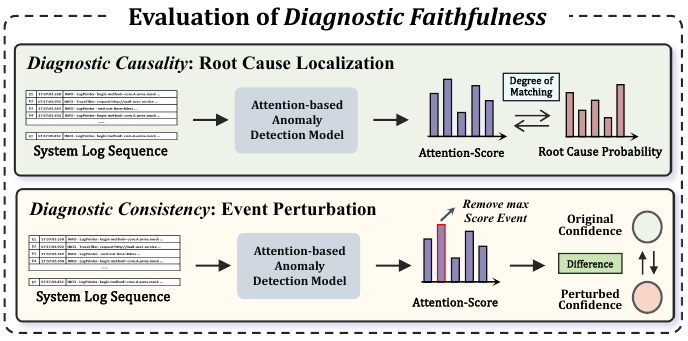}
}
\caption{The proposed attention-based diagnostic faithfulness assessment method for software reliability maintenance systems. These two tasks respectively evaluate the alignment between attention scores and anomaly root causes, and the consistency of between attention scores and detection results.}
\label{teaser}
\vspace{-0.6cm}
\end{figure}

Although these methods have proven effective in detecting system anomalies, their black-box nature inherent to deep learning makes them difficult to deploy in real production environments. Specifically, these deep learning-based anomaly detection systems function as black boxes to service providers utilizing them \cite{poursabzi2021manipulating, zhang2021survey, li2023trustworthy}, allowing providers only to obtain detection results without understanding how anomalies are detected, or whether the results are trustworthy. Providers cannot ascertain whether the anomaly detection system genuinely comprehends the operational patterns of the software system, or is merely overfitting to historical operational states. In other words, these methods intended to maintain software system reliability may themselves be unreliable. Consequently, to avoid unanticipated financial and operational losses, these deep learning-based approaches face significant barriers to deployment in actual production environments.

A comprehensive solution to this problem necessitates opening deep learning's black box, rendering its decision logic fully transparent to service providers, yet substantial research \cite{kaur2022trustworthy, liu2022rethinking, jacovi2020towards, rashkin2021increasing, chan2022comparative} demonstrates this is exceptionally challenging. Therefore, this paper aims to establish a framework for anomaly detection systems to gain service providers' trust, ensuring their trustworthiness when deployed in real production environments. Specifically, through extensive surveys of SREs at a major cloud service provider, we define \textbf{Diagnostic Faithfulness}, which encompasses two critical aspects prioritized by service providers: 1) \textbf{Diagnostic Causality}: whether the anomaly detection system genuinely comprehends the root cause events when anomalies are detected? 2) \textbf{Diagnostic Consistency}: whether the detection results align with the system events the model considers most significant?

To comprehensively assess the diagnostic faithfulness of anomaly detection systems, we designed an attention mechanism-based evaluation method as illustrated in Figure \ref{teaser}. Specifically, it evaluates: 1) the alignment degree between attention scores and root causes when system anomalies are detected; 2) the magnitude of detection result changes when perturbing system events with highest attention scores, to quantify the diagnostic faithfulness of models. Subsequently, we conducted an empirical study on the Thunderbird supercomputer system \cite{oliner2007supercomputers}. Results demonstrate that state-of-the-art existing methods exhibit suboptimal performance in diagnostic faithfulness, rendering their detection outcomes untrustworthy.

To address this issue and enhance the diagnostic faithfulness of anomaly detection systems, we propose \textbf{FaithLog}, a service-provider-faithful log-based failure prediction method. Its core principles are: 1) guiding attention score allocation toward root cause events of anomalies; 2) aligning detection results with system event consistency through adversarial training. Evaluations on two public datasets and one industrial dataset demonstrate that the proposed method remains faithful to service providers and yields trustworthy detection results. 

In summary, our contributions are as follows: 
\begin{itemize}
\item We present the first investigation into the trustworthiness of log-based anomaly detection systems.
\item We define diagnostic faithfulness for log-based anomaly detection systems and establish its evaluation method. 
\item An empirical study revealing that existing log-based anomaly detection systems lack faithfulness. 
\item A service-provider-faithful log-based anomaly detection system, with evaluations confirming its faithfulness.
\end{itemize}

\section{Empirical Study}\label{sec:empirical}
This section evaluates diagnostic faithfulness of existing log-based anomaly detection methods using the Thunderbird system log dataset collected from a supercomputer system \cite{oliner2007supercomputers}. We investigate state-of-the-art attention-based log anomaly detection methods, specifically including LogRobust \cite{zhang2019robust}, PLELog \cite{plelog}, NeuralLog \cite{le2021log}, and SwissLog \cite{li2022swisslog}.

\subsection{Can existing methods comprehend root causes when detecting anomalies?}\label{sec:causality}
This study evaluates the diagnostic causality of existing methods, employing a root cause localization task for assessment. Specifically, we first train baseline methods on anomaly detection tasks then utilize their attention scores for root cause localization. Using Hit Rate (HR@k) as the metric with k=1, 3, 5 for evaluation, the results are presented in the table \ref{tab:empirical_rq1}. Overall, baseline methods achieve only 13.70\% HR@1, 30.42\% HR@3, and 41.13\% HR@5. This indicates that current log-based anomaly methods fail to assign attention scores to root cause events despite detecting system anomalies, rendering model-generated anomaly alerts unrelated to actual root causes, thus making their detection results untrustworthy.

\begin{table}[htbp]
  \centering
  \caption{Diagnostic Causality Evaluation of Existing Methods}
    \begin{tabular}{cccc}
    \midrule
    \multirow{2}[4]{*}{\textbf{Model}} & \multicolumn{3}{c}{\textbf{Metrics}} \\
\cmidrule{2-4}          & \textbf{HR@1} & \textbf{HR@3} & \textbf{HR@5} \\
    \midrule
    LogRobust \cite{zhang2019robust} & 17.19 & 31.83 & 43.14 \\
    PLELog \cite{plelog} & 17.36 & 29.09 & 38.44 \\
    NeuralLog \cite{le2021log} & 13.58 & 39.54 & 53.46 \\
    SwissLog \cite{li2022swisslog} & 6.67  & 21.22 & 29.48 \\
    \midrule
    Average & \textbf{13.70} & \textbf{30.42} & \textbf{41.13} \\
    \midrule
    \end{tabular}%
  \label{tab:empirical_rq1}%
  \vspace{-0.4cm}
\end{table}%

\subsection{Whether the detection results align with the system events the model considers most significant?} \label{sec:consistancy}

This study evaluates diagnostic consistency of existing methods, for which we designed an event perturbation task. Specifically, after model decision-making, we remove the system event with highest attention score and observe confidence changes in new detection results. If confidence decreases, it indicates that the highest-scored event suppresses detection results; conversely, increased confidence suggests support. We employ Support Rate (SR) as the metric, representing the proportion of supportive cases; higher SR indicates better alignment between detection results and most critical system events. Table \ref{tab:empirical_rq2} presents our results, showing an average SR of merely 61.31\%, demonstrating that existing methods do not rely on their identified most critical system events for decision-making, rendering detections untrustworthy. 

\begin{table}[htbp]
  \centering
  \caption{Diagnostic Consistency Evaluation of Existing Methods}
  \scalebox{1}{
    \begin{tabular}{cc}
    \midrule
    \multirow{2}[4]{*}{\textbf{Model}} & \textbf{Metrics} \\
\cmidrule{2-2}          & \textbf{SR} \\
    \midrule
    LogRobust \cite{zhang2019robust} & 71.37 \\
    PLELog \cite{plelog} & 70.69 \\
    NeuralLog \cite{le2021log} & 40.42 \\
    SwissLog \cite{li2022swisslog} & 62.76 \\
    \midrule
    Average & \textbf{61.31} \\
    \midrule
    \end{tabular}%
    }
  \label{tab:empirical_rq2}%
  \vspace{-0.4cm}
\end{table}%

\begin{center}
    \begin{tcolorbox}[colback=gray!10,%gray background
        colframe=black,% black frame colour
        width=\linewidth,% Use 8cm total width,
        arc=1mm, auto outer arc,
        boxrule=0.5pt,
        top=2pt, % Adjust the top space
        bottom=2pt, % Adjust the bottom space
        left=2pt,
        right=2pt
        ]
        \textbf{Summary.} Current log-based anomaly detection methods neither comprehend anomaly root causes during detection nor maintain consistency between results and their prioritized system events, resulting in unreliable outcomes.
    \end{tcolorbox}
\end{center}

\section{Methodology}

\subsection{Overview}
\begin{figure}[t]
\centerline{
\includegraphics[width=1.0\linewidth]{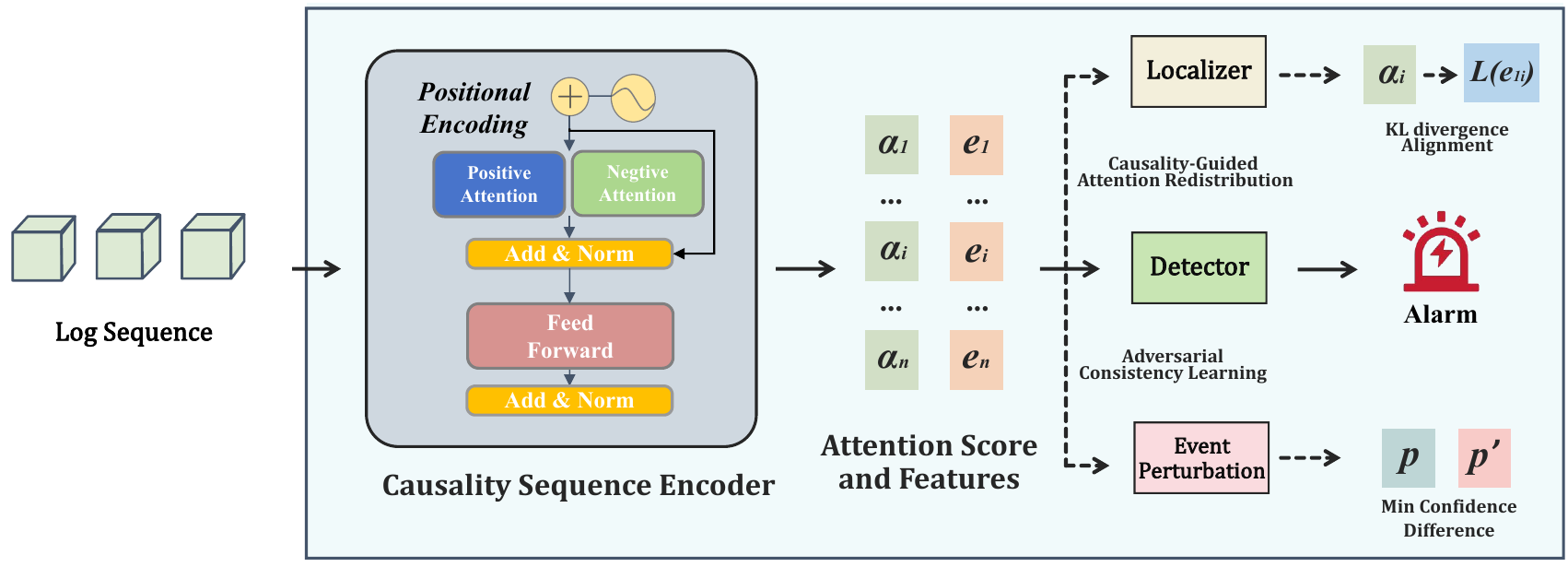}
}
\caption{The overview pipeline of FaithLog.}
\label{pipeline}
\vspace{-0.4cm}
\end{figure}

In this paper, we propose FaithLog, a service-provider-faithful log anomaly detection method. For input log event sequences, FaithLog first employs the advanced log parser Drain \cite{he2017drain} to process unstructured raw logs, to obtain structured log event sequences. Subsequently, following prior work \cite{zhang2024metalog}, we extract semantic embeddings for each log event, which are fed into a Transformer Encoder-based sequence encoder \cite{ma2024knowlog}. Finally, the features encoded by the sequence encoder are delivered to a detector for anomaly identification. During the training phase, FaithLog incorporates a causality-guided attention mechanism and an adversarial consistency learning process, where the former directs attention score allocation toward root cause events of anomalies, while the latter aligns detection results with system event consistency through adversarial learning. Figure \ref{pipeline} illustrates the FaithLog's pipeline.

\subsection{Causality-Guided Attention Mechanism}
Existing log-based anomaly detection methods fail to assign attention scores to root cause events when detecting system anomalies, resulting in model alerts unrelated to actual causes and untrustworthy detection outcomes. To address this issue, FaithLog designs a Causality-Guided Attention Mechanism that aims to allocate maximal attention scores to root cause events, thereby enhancing diagnostic causality.

Specifically, FaithLog first incorporates a negative pathway into the attention mechanism, enabling the emergence of negative attention scores. When assigned negative scores, system events exhibit low probability of being anomaly root causes, yet their magnitude signifies the event's importance, a strategy that enhances the representational capacity of attention scores. Given input log event embeddings $[e_1, e_2, ..., e_n]$, for the $i$-th event embedding $ e_i$, the FaithLog employs a Sinusoidal positional encoding function $PE$ to encode its sequential position vector $p_i$.
\begin{equation}
    \begin{aligned}
PE_{(i,2j)}&=sin(\frac{i}{10000^{\frac{2j}{d_{model}}}}), \\ PE_{(i,2j+1)}&=cos(\frac{i}{10000^{\frac{2j}{d_{model}}}}), 
    \end{aligned}
\end{equation}
Where $d_{model}$ signifies the dimension of $e_i$, with $j$ indexing individual components within $e_i$. The resulting $E = e_i + p_i$, incorporating positional information, is then input to the encoder as the log vector. Inside the encoder, the causality-driven attention mechanism autonomously learns both positive and negative attention weight matrices, calculating attention scores for each log event as the arithmetic difference between positive and negative components.
\begin{equation}
    \begin{aligned}
Q_{pos} &= E W_{pos}^Q, K_{pos} = E W_{pos}^K, V_{pos}=E W_{pos}^V, \\ Q_{neg} &= E W_{neg}^Q, K_{neg} = E W_{neg}^K, V_{neg}=E W_{neg}^V, 
    \end{aligned}
\end{equation}

\begin{equation}
    \begin{aligned}
A = Softmax(\frac{Q_{pos}K_{pos}^T}{\sqrt{d_k}}) V_{pos} - Softmax(\frac{Q_{neg}K_{neg}^T}{\sqrt{d_k}}) V_{neg}, 
    \end{aligned}
\end{equation}

The computed attention scores are utilized to integrate the final features, and optimize the detector by minimizing cross-entropy loss. 
\begin{equation}
    \begin{aligned}
\arg\min_\theta \ \mathcal{L} _{\text{CE}} (\theta) = \sum_{i=1}^{n}[-(y_{i}\log(\hat{y}_{i})+(1-y_{i})\log(1-\hat{y}_{i}))], 
    \end{aligned}
\end{equation}
Subsequently, FaithLog learns root cause scores and guides attention score redistribution. For the input sequence $X = [e_1, e_2, ..., e_n]$, FaithLog establishes a locator $L$ to encode root cause scores $L(e_i)$. Root cause score learning is based on ranking loss \cite{midlog, sultani2018real}. 
\begin{equation}
    \begin{aligned}
\arg\min_\theta \ \mathcal{L} _{\text{Rank}}(\theta) 
=   \max (0, 1 + \max_{e_i \in \mathbf{X}_n} L(e_i) - \max_{e_i \in \mathbf{X}_a} L(e_i) ), 
    \end{aligned}
\end{equation}

Where $X_n$ and $X_a$ represent normal and anomalous runtime log sequences, respectively. Finally, FaithLog aligns attention scores to root cause scores based on KL divergence, to enhance diagnostic causality.
\begin{equation}
    \begin{aligned}
\arg\min_{\theta} \ \mathcal{L} _{\text{KL}}(\theta) = \ KL(L(X)\|A)=\sum_i L(e_i)\log\frac{L(e_i)}{A(e_i)}, 
    \end{aligned}
\end{equation}

\subsection{Adversarial Consistency Learning}
Existing log-based anomaly detection methods exhibit inconsistency between detection outcomes and highest-attention system events during decision-making, resulting in untrustworthy results. To address this issue, FaithLog designs Adversarial Consistency Learning, aiming to align detection results with the most critical system events. 

Specifically, during training, FaithLog first employs detector $D$ for initial detection, and identifies the system event with highest attention score $e_{max} = \arg\max_{i\in{1, 2, ..., n}} a_i$, with initial detection confidence $p = D(X)$. Then, FaithLog removes the highest-attention system event $e_{max}$, and performs a secondary detection, yielding new confidence $p' = D(X\setminus e_{max})$. FaithLog establishes the following constraint during training, which encourages reduced detection confidence after removing the most critical system event, thereby enhancing diagnostic consistency.
\begin{equation}
    \begin{aligned}
\arg\min_{\theta} \ \mathcal{L} _{\text{Consistency}}(\theta) = \max(1 + p' - p) , 
    \end{aligned}
\end{equation}

\subsection{Training}
Following this approach, FaithLog can be deployed on real industrial software systems, to deliver reliability maintenance strategies faithful to service providers. FaithLog undergoes end-to-end training by integrating the aforementioned loss functions:
\begin{equation}
    \begin{aligned}
\arg\min_{\theta} \ \mathcal{L} (\theta) = \lambda_1 \mathcal{L} _{\text{CE}} (\theta) + \lambda_2 \mathcal{L} _{\text{Rank}} + \lambda_3 \mathcal{L} _{\text{KL}}+ \lambda_4 \mathcal{L} _{\text{Consistency}} , 
    \end{aligned}
\end{equation}
where $\lambda_1, \lambda_2, \lambda_3, \lambda_4$ are hyperparameters for balancing the loss functions.

\begin{table*}[htbp]
  \centering
  \caption{Diagnostic Causality Evaluation of FaithLog}
  \scalebox{0.8}{
    \begin{tabular}{rrrrrrrrrr}
    \midrule
    \multicolumn{1}{c}{\multirow{2}[4]{*}{\textbf{Dataset}}} & \multicolumn{1}{c}{\multirow{2}[4]{*}{\textbf{Model}}} & \multicolumn{8}{c}{\textbf{Metrics}} \\
\cmidrule{3-10}          &       & \multicolumn{1}{c}{\textbf{HR@1}} & \multicolumn{1}{c}{\textbf{HR@3}} & \multicolumn{1}{c}{\textbf{HR@5}} & \multicolumn{1}{c}{\textbf{PR@3}} & \multicolumn{1}{c}{\textbf{PR@5}} & \multicolumn{1}{c}{\textbf{MAP@3}} & \multicolumn{1}{c}{\textbf{MAP@5}} & \multicolumn{1}{c}{\textbf{MRR}} \\
    \midrule
    \multicolumn{1}{c}{\multirow{5}[4]{*}{BGL}} & \multicolumn{1}{c}{LogRobust \cite{zhang2019robust}} & \multicolumn{1}{c}{37.98} & \multicolumn{1}{c}{65.13} & \multicolumn{1}{c}{67.89} & \multicolumn{1}{c}{39.90} & \multicolumn{1}{c}{41.15} & \multicolumn{1}{c}{39.26} & \multicolumn{1}{c}{39.94} & \multicolumn{1}{c}{52.78} \\
          & \multicolumn{1}{c}{PLELog \cite{plelog}} & \multicolumn{1}{c}{44.32} & \multicolumn{1}{c}{65.30} & \multicolumn{1}{c}{73.52} & \multicolumn{1}{c}{47.59} & \multicolumn{1}{c}{51.27} & \multicolumn{1}{c}{46.07} & \multicolumn{1}{c}{51.27} & \multicolumn{1}{c}{58.22} \\
          & \multicolumn{1}{c}{NeuralLog \cite{le2021log}} & \multicolumn{1}{c}{37.04} & \multicolumn{1}{c}{49.83} & \multicolumn{1}{c}{52.28} & \multicolumn{1}{c}{40.87} & \multicolumn{1}{c}{42.69} & \multicolumn{1}{c}{39.09} & \multicolumn{1}{c}{40.38} & \multicolumn{1}{c}{43.49} \\
          & \multicolumn{1}{c}{SwissLog \cite{li2022swisslog}} & \multicolumn{1}{c}{31.89} & \multicolumn{1}{c}{37.04} & \multicolumn{1}{c}{39.24} & \multicolumn{1}{c}{31.55} & \multicolumn{1}{c}{31.90} & \multicolumn{1}{c}{31.98} & \multicolumn{1}{c}{32.28} & \multicolumn{1}{c}{36.84} \\
\cmidrule{2-10}          & \multicolumn{1}{c}{FaithLog} & \multicolumn{1}{c}{\textbf{70.03}} & \multicolumn{1}{c}{\textbf{77.29}} & \multicolumn{1}{c}{\textbf{79.42}} & \multicolumn{1}{c}{\textbf{72.56}} & \multicolumn{1}{c}{\textbf{73.70}} & \multicolumn{1}{c}{\textbf{71.31}} & \multicolumn{1}{c}{\textbf{72.18}} & \multicolumn{1}{c}{\textbf{74.29}} \\
    \midrule
    \multicolumn{1}{c}{\multirow{5}[4]{*}{Thunderbird}} & \multicolumn{1}{c}{LogRobust \cite{zhang2019robust}} & \multicolumn{1}{c}{17.19} & \multicolumn{1}{c}{31.83} & \multicolumn{1}{c}{43.14} & \multicolumn{1}{c}{24.11} & \multicolumn{1}{c}{32.24} & \multicolumn{1}{c}{20.53} & \multicolumn{1}{c}{24.40} & \multicolumn{1}{c}{20.76} \\
          & \multicolumn{1}{c}{PLELog \cite{plelog}} & \multicolumn{1}{c}{17.36} & \multicolumn{1}{c}{29.09} & \multicolumn{1}{c}{38.44} & \multicolumn{1}{c}{23.22} & \multicolumn{1}{c}{30.07} & \multicolumn{1}{c}{20.29} & \multicolumn{1}{c}{23.47} & \multicolumn{1}{c}{29.66} \\
          & \multicolumn{1}{c}{NeuralLog \cite{le2021log}} & \multicolumn{1}{c}{13.58} & \multicolumn{1}{c}{39.54} & \multicolumn{1}{c}{53.46} & \multicolumn{1}{c}{31.14} & \multicolumn{1}{c}{43.11} & \multicolumn{1}{c}{22.77} & \multicolumn{1}{c}{29.86} & \multicolumn{1}{c}{30.33} \\
          & \multicolumn{1}{c}{SwissLog \cite{li2022swisslog}} & \multicolumn{1}{c}{6.67} & \multicolumn{1}{c}{21.22} & \multicolumn{1}{c}{29.48} & \multicolumn{1}{c}{14.36} & \multicolumn{1}{c}{19.89} & \multicolumn{1}{c}{10.78} & \multicolumn{1}{c}{13.90} & \multicolumn{1}{c}{17.93} \\
\cmidrule{2-10}          & \multicolumn{1}{c}{FaithLog} & \multicolumn{1}{c}{\textbf{34.13}} & \multicolumn{1}{c}{\textbf{48.93}} & \multicolumn{1}{c}{\textbf{55.86}} & \multicolumn{1}{c}{\textbf{40.82}} & \multicolumn{1}{c}{\textbf{47.84}} & \multicolumn{1}{c}{\textbf{37.54}} & \multicolumn{1}{c}{\textbf{40.97}} & \multicolumn{1}{c}{\textbf{45.86}} \\
    \midrule
    \multicolumn{1}{c}{\multirow{5}[4]{*}{System A}} & \multicolumn{1}{c}{LogRobust \cite{zhang2019robust}} & \multicolumn{1}{c}{44.15} & \multicolumn{1}{c}{54.24} & \multicolumn{1}{c}{60.58} & \multicolumn{1}{c}{45.28} & \multicolumn{1}{c}{47.50} & \multicolumn{1}{c}{44.70} & \multicolumn{1}{c}{45.56} & \multicolumn{1}{c}{53.40} \\
          & \multicolumn{1}{c}{PLELog \cite{plelog}} & \multicolumn{1}{c}{52.41} & \multicolumn{1}{c}{63.13} & \multicolumn{1}{c}{67.93} & \multicolumn{1}{c}{53.29} & \multicolumn{1}{c}{54.84} & \multicolumn{1}{c}{52.80} & \multicolumn{1}{c}{53.46} & \multicolumn{1}{c}{59.14} \\
          & \multicolumn{1}{c}{NeuralLog \cite{le2021log}} & \multicolumn{1}{c}{52.45} & \multicolumn{1}{c}{67.19} & \multicolumn{1}{c}{74.72} & \multicolumn{1}{c}{53.71} & \multicolumn{1}{c}{56.32} & \multicolumn{1}{c}{52.93} & \multicolumn{1}{c}{54.05} & \multicolumn{1}{c}{61.64} \\
          & \multicolumn{1}{c}{SwissLog \cite{li2022swisslog}} & \multicolumn{1}{c}{43.57} & \multicolumn{1}{c}{54.14} & \multicolumn{1}{c}{56.90} & \multicolumn{1}{c}{45.20} & \multicolumn{1}{c}{47.17} & \multicolumn{1}{c}{44.34} & \multicolumn{1}{c}{45.28} & \multicolumn{1}{c}{49.41} \\
\cmidrule{2-10}          & \multicolumn{1}{c}{FaithLog} & \multicolumn{1}{c}{\textbf{58.41}} & \multicolumn{1}{c}{\textbf{76.73}} & \multicolumn{1}{c}{\textbf{81.63}} & \multicolumn{1}{c}{\textbf{62.13}} & \multicolumn{1}{c}{\textbf{65.06}} & \multicolumn{1}{c}{\textbf{60.30}} & \multicolumn{1}{c}{\textbf{61.95}} & \multicolumn{1}{c}{\textbf{68.36}} \\
    \midrule
    \end{tabular}%
    }
  \label{tab:rq1}%
  \vspace{-0.6cm}
\end{table*}%

\begin{table}[htbp]
  \centering
  \caption{Diagnostic Consistency Evaluation of FaithLog}
  \scalebox{0.9}{
    \begin{tabular}{rrr}
    \midrule
    \multicolumn{1}{c}{\multirow{2}[4]{*}{\textbf{Dataset}}} & \multicolumn{1}{c}{\multirow{2}[4]{*}{\textbf{Model}}} & \multicolumn{1}{c}{\textbf{Metrics}} \\
\cmidrule{3-3}          &       & \multicolumn{1}{c}{\textbf{SR}} \\
    \midrule
    \multicolumn{1}{c}{\multirow{5}[4]{*}{BGL}} & \multicolumn{1}{c}{LogRobust \cite{zhang2019robust}} & \multicolumn{1}{c}{77.86} \\
          & \multicolumn{1}{c}{PLELog \cite{plelog}} & \multicolumn{1}{c}{81.76} \\
          & \multicolumn{1}{c}{NeuralLog \cite{le2021log}} & \multicolumn{1}{c}{66.44} \\
          & \multicolumn{1}{c}{SwissLog \cite{li2022swisslog}} & \multicolumn{1}{c}{54.58} \\
\cmidrule{2-3}          & \multicolumn{1}{c}{FaithLog} & \multicolumn{1}{c}{\textbf{88.58}} \\
    \midrule
    \multicolumn{1}{c}{\multirow{5}[4]{*}{Thunderbird}} & \multicolumn{1}{c}{LogRobust \cite{zhang2019robust}} & \multicolumn{1}{c}{71.37} \\
          & \multicolumn{1}{c}{PLELog \cite{plelog}} & \multicolumn{1}{c}{70.69} \\
          & \multicolumn{1}{c}{NeuralLog \cite{le2021log}} & \multicolumn{1}{c}{40.42} \\
          & \multicolumn{1}{c}{SwissLog \cite{li2022swisslog}} & \multicolumn{1}{c}{62.76} \\
\cmidrule{2-3}          & \multicolumn{1}{c}{FaithLog} & \multicolumn{1}{c}{\textbf{82.74}} \\
    \midrule
    \multicolumn{1}{c}{\multirow{5}[4]{*}{System A}} & \multicolumn{1}{c}{LogRobust \cite{zhang2019robust}} & \multicolumn{1}{c}{75.73} \\
          & \multicolumn{1}{c}{PLELog \cite{plelog}} & \multicolumn{1}{c}{52.73} \\
          & \multicolumn{1}{c}{NeuralLog \cite{le2021log}} & \multicolumn{1}{c}{48.07} \\
          & \multicolumn{1}{c}{SwissLog \cite{li2022swisslog}} & \multicolumn{1}{c}{66.34} \\
\cmidrule{2-3}          & \multicolumn{1}{c}{FaithLog} & \multicolumn{1}{c}{\textbf{83.28}} \\
    \midrule
    \end{tabular}%
    }
  \label{tab:rq2}%
  \vspace{-0.4cm}
\end{table}%

\section{Preliminary Evaluation}

\subsection{Experimental Setup}
We conducted experiments on two public datasets (BGL and Thunderbird) \cite{oliner2007supercomputers} and an industrial log dataset System A collected from real-world systems, to assess FaithLog's diagnostic faithfulness. Detailed experimental configurations are provided in Section \ref{sec:empirical}. 
% For a more comprehensive assessment, in the root cause localization task, we employ Hit Rate at K (HR@k), Precision at K (PR@k), Mean Average Precision at K (MAP@k), and Mean Reciprocal Rank at K (MRR) for evaluation.
For a more comprehensive assessment, in the root cause localization task, we employ HR@k, PR@k, MAP@k, and MRR for evaluation \cite{zheng2024lemma}.

\subsection{Diagnostic Causality of FaithLog}
% This section evaluates whether FaithLog comprehends root causes when detecting anomalies, for which we follow the Section \ref{sec:causality} configuration and utilize attention scores for root cause localization, with Table \ref{tab:rq1} presenting comparative results against existing methods. 

This section evaluates whether FaithLog comprehends root causes when detecting anomalies, with Table \ref{tab:rq1} presenting comparative results against existing methods. Overall, FaithLog achieves superior performance across all three evaluation datasets. Existing methods employ naive attention mechanisms, where attention scores only indicate importance without representing causality, potentially assigning high scores even to normal system events. Conversely, FaithLog incorporates a carefully designed causality-guided attention mechanism, assigning maximally possible attention scores to root cause events, effectively enhancing diagnostic causality. These findings demonstrate FaithLog's substantial potential for direct application in root cause localization.

\subsection{Diagnostic Consistency of FaithLog}
% This section evaluates whether FaithLog's detection results maintain consistency with the system events it deems most critical. For this purpose, we follow the setting in Section \ref{sec:consistancy} and conduct evaluation through event perturbation, with Table \ref{tab:rq2} presenting comparative results. 
This section evaluates whether FaithLog's detection results maintain consistency with the system events it deems most critical, with Table \ref{tab:rq2} presenting results. 
Notably, SR of existing methods are suboptimal across all evaluated datasets. This occurs because existing methods merely utilize attention to combine features of system events, where event weights fail to represent their capacity to influence detection, resulting in attention scores being incapable of explaining the model's decision logic. Conversely, FaithLog benefits from meticulously designed adversarial consistency learning, maintaining high consistency between detection results and its prioritized system events, thereby enabling attention scores to reflect the model's decision logic, consequently empowering it to deliver trustworthy anomaly detection results for service providers.

\section{Conclusion and Future Work}
In this paper, we present the first study on the trustworthiness of software reliability maintenance systems. 
Through surveys of SREs at a major cloud service provider, we designed a metric—diagnostic faithfulness—for evaluating the trustworthiness of software reliability maintenance systems, and developed two evaluation tasks: attention-based root cause localization and event perturbation. 
Empirical studies demonstrate that existing methods fail to comprehend root causes when detecting system anomalies, and exhibit inconsistency between detection outcomes and their prioritized system events, rendering their detection results untrustworthy. 
Therefore, we propose FaithLog, a faithful log-based anomaly detection system. This work provides a pathway for assessing the trustworthiness of maintenance systems, and we plan to deliver transparent, trustworthy, and explainable reliability maintenance systems in future work.

\section*{Acknowledgment}
This work was supported by National Key Laboratory of Data Space Technology and System.

\bibliographystyle{IEEEtran}
\bibliography{sample}

\end{document}